\documentclass[%
superscriptaddress,
 preprint,
 amsmath,amssymb,
 aps,
pra,
floatfix,
]{revtex4-2}

\usepackage[utf8]{inputenc}
\usepackage{graphicx}
\usepackage{booktabs}
\usepackage{siunitx} 
\usepackage{chemformula}
\let\ce\ch
\usepackage{longtable}
\usepackage[caption=false]{subfig}
\usepackage{hyperref}
\usepackage{comment}

\begin{document}

\title{International Optical Clock Comparison Using the European Optical Fiber Network}



\author{Marco~Pizzocaro}
\email[]{m.pizzocaro@inrim.it}
\affiliation{Istituto Nazionale di Ricerca Metrologica (INRIM), Strada delle Cacce 91, 10135, Torino, Italy}

\author{Clara~Zyskind}
\affiliation{Laboratoire Temps Espace (LNE-OP), Observatoire de Paris, Université PSL, Sorbonne Université, Université de Lille, LNE, CNRS, 61 avenue de l'Observatoire, 75014 Paris, France}

\author{Anne~Amy-Klein}
\affiliation{Laboratoire de Physique des Lasers (LPL), Université Sorbonne Paris Nord, CNRS, 99 Avenue Jean-Baptiste Clément, 93430 Villetaneuse, France.}

\author{Erik~Benkler}
\affiliation{Physikalisch-Technische Bundesanstalt (PTB), Bundesallee 100, 38116 Braunschweig, Germany}

\author{Sebastien~Bize}
\affiliation{Laboratoire Temps Espace (LNE-OP), Observatoire de Paris, Université PSL, Sorbonne Université, Université de Lille, LNE, CNRS, 61 avenue de l'Observatoire, 75014 Paris, France}

\author{Davide~Calonico}
\affiliation{Istituto Nazionale di Ricerca Metrologica (INRIM), Strada delle Cacce 91, 10135, Torino, Italy}

\author{Etienne~Cantin}
\affiliation{Laboratoire de Physique des Lasers (LPL), Université Sorbonne Paris Nord, CNRS, 99 Avenue Jean-Baptiste Clément, 93430 Villetaneuse, France.}

\author{Christian Chardonnet}
\affiliation{Laboratoire de Physique des Lasers (LPL), Université Sorbonne Paris Nord, CNRS, 99 Avenue Jean-Baptiste Clément, 93430 Villetaneuse, France.}

\author{Cecilia~Clivati}
\affiliation{Istituto Nazionale di Ricerca Metrologica (INRIM), Strada delle Cacce 91, 10135, Torino, Italy}

\author{Stefano~Condio}
\affiliation{Istituto Nazionale di Ricerca Metrologica (INRIM), Strada delle Cacce 91, 10135, Torino, Italy}

\author{E. Anne~Curtis}
\affiliation{National Physical Laboratory (NPL), Hampton Road, Teddington, TW11 0LW, UK }

\author{Simone~Donadello}
\affiliation{Istituto Nazionale di Ricerca Metrologica (INRIM), Strada delle Cacce 91, 10135, Torino, Italy}

\author{Sören~Dörscher}
\affiliation{Physikalisch-Technische Bundesanstalt (PTB), Bundesallee 100, 38116 Braunschweig, Germany}

\author{Chen-Hao~Feng}
\affiliation{National Physical Laboratory (NPL), Hampton Road, Teddington, TW11 0LW, UK }

\author{Melina~Filzinger}
\affiliation{Physikalisch-Technische Bundesanstalt (PTB), Bundesallee 100, 38116 Braunschweig, Germany}

\author{Jacques-Olivier~Gaudron}
\affiliation{National Physical Laboratory (NPL), Hampton Road, Teddington, TW11 0LW, UK }

\author{Rachel~M.~Godun}
\affiliation{National Physical Laboratory (NPL), Hampton Road, Teddington, TW11 0LW, UK }

\author{Irene~Goti}
\affiliation{Istituto Nazionale di Ricerca Metrologica (INRIM), Strada delle Cacce 91, 10135, Torino, Italy}

\author{Ian~R.~Hill}
\affiliation{National Physical Laboratory (NPL), Hampton Road, Teddington, TW11 0LW, UK }

\author{Wei~Huang}
\affiliation{National Physical Laboratory (NPL), Hampton Road, Teddington, TW11 0LW, UK }

\author{Nils~Huntemann}
\affiliation{Physikalisch-Technische Bundesanstalt (PTB), Bundesallee 100, 38116 Braunschweig, Germany}

\author{Matthew	Johnson}
\affiliation{National Physical Laboratory (NPL), Hampton Road, Teddington, TW11 0LW, UK }

\author{Joshua~Klose}
\affiliation{Physikalisch-Technische Bundesanstalt (PTB), Bundesallee 100, 38116 Braunschweig, Germany}

\author{Jochen~Kronjäger}
\affiliation{Physikalisch-Technische Bundesanstalt (PTB), Bundesallee 100, 38116 Braunschweig, Germany}

\author{Alexander~Kuhl}
\affiliation{Physikalisch-Technische Bundesanstalt (PTB), Bundesallee 100, 38116 Braunschweig, Germany}

\author{Rodolphe~Le Targat}
\affiliation{Laboratoire Temps Espace (LNE-OP), Observatoire de Paris, Université PSL, Sorbonne Université, Université de Lille, LNE, CNRS, 61 avenue de l'Observatoire, 75014 Paris, France}

\author{Filippo~Levi}
\affiliation{Istituto Nazionale di Ricerca Metrologica (INRIM), Strada delle Cacce 91, 10135, Torino, Italy}

\author{Burghard~Lipphardt}
\affiliation{Physikalisch-Technische Bundesanstalt (PTB), Bundesallee 100, 38116 Braunschweig, Germany}

\author{Christian~Lisdat}
\affiliation{Physikalisch-Technische Bundesanstalt (PTB), Bundesallee 100, 38116 Braunschweig, Germany}

\author{Jerome~Lodewyck}
\affiliation{Laboratoire Temps Espace (LNE-OP), Observatoire de Paris, Université PSL, Sorbonne Université, Université de Lille, LNE, CNRS, 61 avenue de l'Observatoire, 75014 Paris, France}

\author{Olivier~Lopez}
\affiliation{Laboratoire de Physique des Lasers (LPL), Université Sorbonne Paris Nord, CNRS, 99 Avenue Jean-Baptiste Clément, 93430 Villetaneuse, France.}

\author{Helen~S.~Margolis}
\affiliation{National Physical Laboratory (NPL), Hampton Road, Teddington, TW11 0LW, UK }

\author{Maxime~Mazouth-Laurol}
\affiliation{Laboratoire Temps Espace (LNE-OP), Observatoire de Paris, Université PSL, Sorbonne Université, Université de Lille, LNE, CNRS, 61 avenue de l'Observatoire, 75014 Paris, France}

\author{Alberto~Mura}
\affiliation{Istituto Nazionale di Ricerca Metrologica (INRIM), Strada delle Cacce 91, 10135, Torino, Italy}

\author{Benjamin~Pointard}
\affiliation{Laboratoire Temps Espace (LNE-OP), Observatoire de Paris, Université PSL, Sorbonne Université, Université de Lille, LNE, CNRS, 61 avenue de l'Observatoire, 75014 Paris, France}

\author{Paul-Eric~Pottie}
\affiliation{Laboratoire Temps Espace (LNE-OP), Observatoire de Paris, Université PSL, Sorbonne Université, Université de Lille, LNE, CNRS, 61 avenue de l'Observatoire, 75014 Paris, France}

\author{Matias~Risaro}
\affiliation{Istituto Nazionale di Ricerca Metrologica (INRIM), Strada delle Cacce 91, 10135, Torino, Italy}
\affiliation{National Physical Laboratory (NPL), Hampton Road, Teddington, TW11 0LW, UK }

\author{Billy~I.~Robertson}
\affiliation{National Physical Laboratory (NPL), Hampton Road, Teddington, TW11 0LW, UK }

\author{Marco~Schioppo}
\affiliation{National Physical Laboratory (NPL), Hampton Road, Teddington, TW11 0LW, UK }

\author{Kilian~Stahl}
\affiliation{Physikalisch-Technische Bundesanstalt (PTB), Bundesallee 100, 38116 Braunschweig, Germany}

\author{Martin~Steinel}
\affiliation{Physikalisch-Technische Bundesanstalt (PTB), Bundesallee 100, 38116 Braunschweig, Germany}

\author{Alexandra~Tofful}
\affiliation{National Physical Laboratory (NPL), Hampton Road, Teddington, TW11 0LW, UK }

\author{Mads~Tønnes}
\affiliation{Laboratoire Temps Espace (LNE-OP), Observatoire de Paris, Université PSL, Sorbonne Université, Université de Lille, LNE, CNRS, 61 avenue de l'Observatoire, 75014 Paris, France}

\author{Jacob~Tunesi}
\affiliation{National Physical Laboratory (NPL), Hampton Road, Teddington, TW11 0LW, UK }

\date{\today}

\begin{abstract}

Optical clocks have achieved remarkable estimated  fractional frequency uncertainties reaching the \num{e-18} level and below, enabling applications in fundamental physics, general relativity, and geodesy. However, the challenge of verifying the international consistency of optical clocks remains critical as efforts intensify toward redefining the SI second based on an optical transition or transitions. We report on a two-month international clock comparison campaign involving seven optical clocks in four national metrology institutes (INRIM, LNE-OP, NPL, and PTB) connected via the optical fiber network established in Europe. The campaign resulted in optical frequency ratios with uncertainties ranging from \num{7.7e-18} to \num{6.1e-17}. Among the results, the \ce{^{171}Yb+}(E3) clocks at NPL and PTB demonstrated agreement within an uncertainty of \num{7.7e-18}, marking the first international verification of two independently developed optical clocks below one part in \num{e17}. 
The operation of the \ce{^{199}Hg} clock at LNE-OP (formerly LNE-SYRTE)  resulted in frequency ratios with improved uncertainties with  \ce{^{171}Yb+}(E3), \ce{^{171}Yb}, and \ce{^{87}Sr} optical clocks.
These results provide input for the redefinition of the second and underscore how fiber-linked clock networks can advance metrology and scientific applications.  
\end{abstract}  

\maketitle

\section{Introduction}
Optical clocks are at the forefront of precision measurement, achieving fractional frequency uncertainties below the \num{3e-18} level \cite{McGrew2018, Sanner2019, Huang2022, Aeppli2024,  Ma2024, Tofful2024, Hausser2025, Marshall2025, Lindvall2025a}, and demonstrating unprecedented instabilities below \num{6e-17} at 1 s \cite{Oelker2019, Schioppo2017}. These advances have enabled groundbreaking applications, including tests of fundamental physics such as searching for variations of fundamental constants \cite{Lange2021, Sherrill2023}, the testing of general relativity \cite{Takamoto2020, Delva2017}, chronometric leveling \cite{Mehlstaeubler2018, Grotti2018}, searches for dark matter \cite{Filzinger2025,Filzinger2023,Kobayashi2022,Roberts2020, Beloy2021} and the generation of optical time scales \cite{Formichella2024,Yao2019,Hachisu2018}.
On the one hand, the consistency of optical clocks has been demonstrated with an uncertainty at the \num{5e-18} level and below
within single laboratories operating clocks based on the same atomic species \cite{Sanner2019,McGrew2018,Ushijima2015,Zhigiang2023} . On the other hand, frequency ratio measurements between clocks based on different atomic species have been achieved both locally \cite{Dorscher2021,Ohmae2020, Ohtsubo2020, Nemitz2016,Rosenband2008,Hausser2025} and remotely via optical fiber links \cite{Clivati2022a, Beloy2021},  satellites \cite{Riedel2020,Fujieda2018,Hachisu2014}, very-long-baseline interferometry \cite{Pizzocaro2021, Negusini2026} or  contributions of optical clocks to international time scales \cite{Pizzocaro2020,McGrew2019}.  

These advances have prompted the international scientific community to develop a roadmap toward the redefinition of the second in the International System of Units (SI) based on optical clocks \cite{Dimarcq2024}. This roadmap sets several criteria which have yet to be fully met. Among these, the criterion of verifying the international consistency of optical clocks through independent comparisons remains a key challenge, necessitating further efforts and measurement campaigns.

Measuring an optical frequency ratio in an international clock comparison campaign is particularly demanding, as clocks must operate reliably within a limited timeframe and achieve their expected performance on a predetermined schedule. It also requires long-distance fiber links with high uptime, often with multiple relay stations to extend their reach, as well as optical frequency combs at each laboratory. The different laboratories must coordinate their measurements, ensuring that all segments of the comparison chain function correctly and simultaneously, while also ensuring that a consistent data recording formalism is followed by the participants \cite{Lodewyck2020}. This process is inherently more challenging than optical clock comparisons within the same laboratory.

\begin{figure}
    \includegraphics[width=0.7\textwidth]{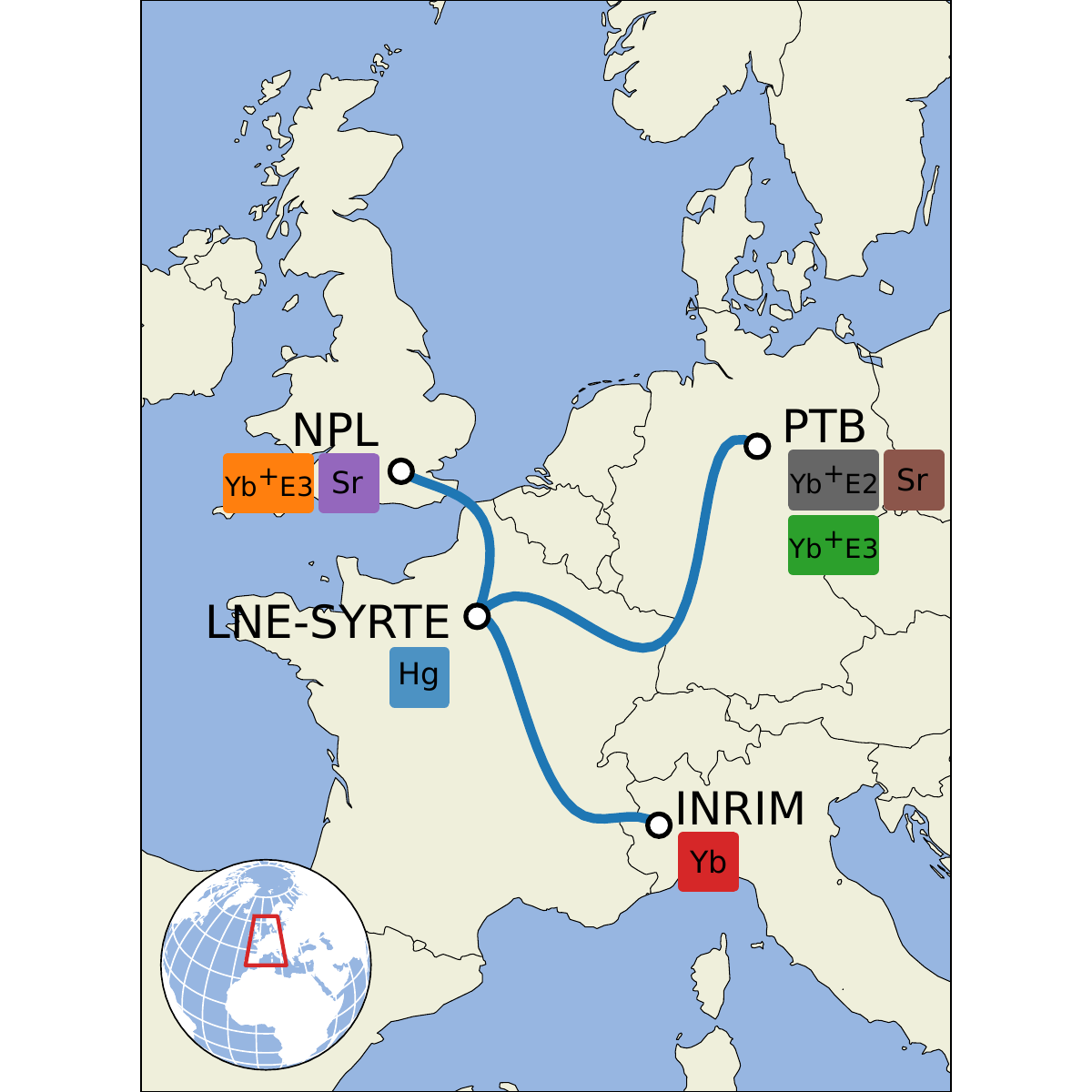}
    \caption{Map of the clock comparison, showing the location of the optical clocks and of the optical fiber links.}
    \label{fig:map}
\end{figure}

In this work we report on a clock comparison campaign involving 7 optical clocks in 4 different countries connected by optical fiber links carried out in early 2023. A map of the fiber links and of the clocks is shown in Figure \ref{fig:map}. The clocks involved are the \ce{^{199}Hg} optical lattice clock at LNE-OP (formerly LNE-SYRTE \footnote{We refer to the clock as it was designated at the time of the measurements to remain consistent with the nomenclature used in our lab and in other collaborations. Therefore, the clocks and comparison campaigns conducted at LNE-OP are labeled with the ``SYRTE'' prefix.}) in France (SYRTE-Hg), the \ce{^{171}Yb+} ion optical clock operating on the electric octupole (E3) transition at NPL in the UK (NPL-E3Yb+3), the \ce{^{171}Yb+} ion optical clocks operating on both the E3 and electric quadrupole (E2) transitions at PTB in Germany (PTB-Yb1E3 and PTB-Yb1E2), the \ce{^{171}Yb} optical lattice clock at INRIM in Italy (IT-Yb1) and the \ce{^{87}Sr} optical lattice clocks at NPL and PTB (NPL-Sr1 and PTB-Sr3). The comparison uses the European network of phase-stabilized fiber links 
connecting LNE-OP and NPL (785 km), LNE-OP and PTB (1370 km), and LNE-OP and INRIM (1023 km) \cite{Clivati2022a,Schioppo2022,Cantin2021,Lisdat2016}. The campaign lasted 65 days from the 24th of February 2023 (Modified Julian Date, MJD, 59999).

This work builds upon two other clock comparison campaigns which shared some of the clocks and parts of the fiber infrastructure.
First, a large coordinated comparison campaign was carried out in March 2022 as part of the European project ROCIT \cite{Rocit2025}. The ROCIT campaign involved 10 optical clocks at INRIM, NPL, PTB, SYRTE, as well as VTT in Finland and NMIJ in Japan. All clocks were compared using the Global Navigation Satellite Systems (GNSS). Clocks at INRIM, PTB and SYRTE were also compared using the European fiber link network. Clocks at INRIM, NPL, and PTB participated in both the 2023 campaign and the 2022 ROCIT campaign. SYRTE operated their \ce{^{87}Sr} optical lattice clock in 2022 and their \ce{^{199}Hg} optical lattice clock in 2023. 

Another campaign  was also running in March and April 2023 for the project ICON \cite{Icon2024} involving NPL-Sr1 and PTB-Sr3 as well as the transportable \ce{^{87}Sr} optical lattice clocks developed at PTB \cite{Nosske25} and RIKEN in Japan \cite{Ohmae2021,Takamoto2020}. NPL-Sr1 and PTB-Sr3 contributed at the same time for the ICON campaign and the measurements presented in this paper are based on the same dataset.

\section{Results}
\begin{figure}
    \subfloat[]{\includegraphics[width=0.5\textwidth]{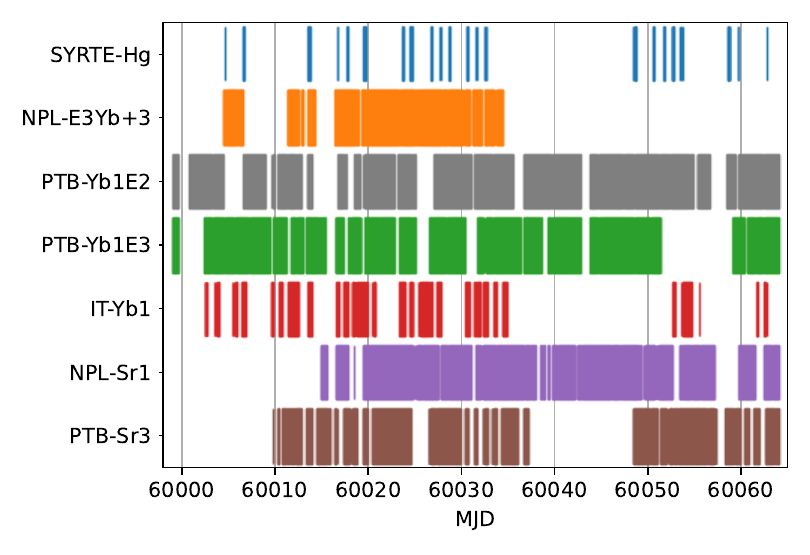}}
    \subfloat[]{\includegraphics[width=0.5\textwidth]{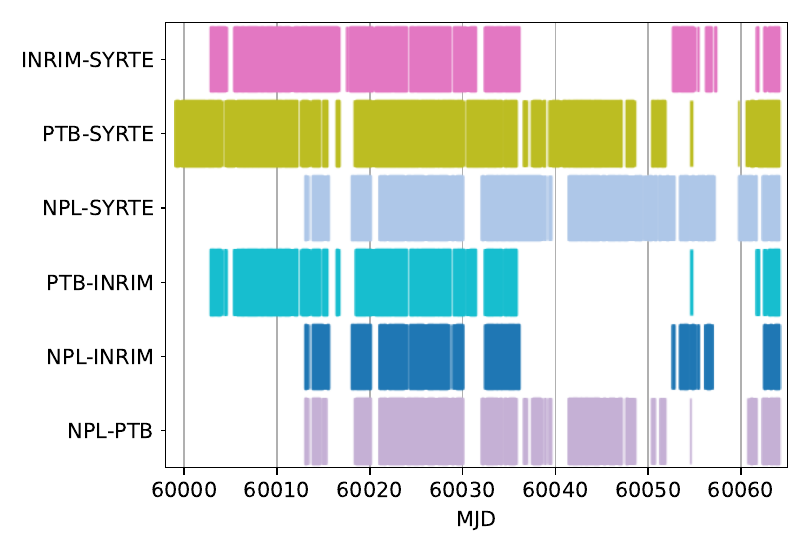}}
    \caption{Uptimes of a) clocks and b) fiber links during the comparison campaign. Colored regions mark the times with valid data. The date is reported as Modified Julian Date (MJD), starting from MJD 59999 (24th February 2023).}
    \label{fig:Uptimes}
\end{figure}

An overview of the clocks participating in the campaign is shown in Table \ref{tab:clocks}.
References within the table provide details on the operation of each clock.
Figure \ref{fig:Uptimes} shows the uptimes—defined as the periods with valid data—of the clocks and fiber links during the campaign.

Each institute was responsible for their local clocks, their fiber link equipment, and their optical frequency combs required for the comparison. In France, fiber links of the REFIMEVE network are operated by LPL and LNE-OP.

For clarity and to avoid redundancy, not all possible combinations of frequency ratios between clocks in the network are explicitly reported.

The PTB Yb ion clocks (PTB-Yb1E2 and PTB-Yb1E3) share the same physics package and were operated at the same time in an interleaved manner. Given that the uptimes of the two clocks were similar, we report only the local \ce{^{171}Yb+}(E2)/\ce{^{171}Yb+}(E3) frequency ratio, rather than all combinations of PTB-Yb1E2 with other clocks. Frequency ratios between other clocks in the network can be obtained by combining the presented ratios, with the corresponding uncertainty evaluated using the correlation matrix.

The PTB Sr lattice clock (PTB-Sr3) was subject to a frequency shift caused by the clock laser \cite{Klose2026}, which likely reduced its frequency by several $10^{-17}$ and cannot be corrected retrospectively \cite{Icon2024}. Consequently, the values of frequency ratios involving PTB-Sr3 are not reported in this work. We do not apply a correction similar to that introduced in Ref.~\cite{Hausser2025}, as such a treatment would not add independent constraints beyond those already provided by measured frequency ratios and the data presented here.  We nevertheless report the measured instability of ratios involving PTB-Sr3, as this clock exhibited the lowest instability within the network and therefore provided an assessment of the short-term performance of the other clocks. The identified frequency shift is not expected to have significantly affected the observed instability of PTB-Sr3.

\begin{table}
\centering
\caption{Clocks participating in the measurement campaign. For each clock, the table reports its name, the clock species, the fractional systematic uncertainty $u_\text{B}$, and the uncertainty of the relativistic redshift correction $u_\text{RRS}$. This last uncertainty applies only for remote comparisons, as for the local comparisons the uncertainty is smaller (\num{1.2e-18} at NPL and negligible for PTB-Yb1E2/PTB-Yb1E3 as the two clocks share the same trap). PTB-Sr3 was found to be affected by a frequency shift that cannot be corrected retrospectively.}
\small
\begin{tabular}{llSSc} 
\toprule
Clock   & Clock species & {$u_\text{B}$}  & {$u_\text{RRS}$}    & Ref. \\ 
\midrule
SYRTE-Hg    & \ce{^{199}Hg} & 5.4e-17   & 3.0e-18 & \cite{Tyumenev2016}\\
PTB-Yb1E2     & \ce{^{171}Yb+}(E2) & 2.6e-17   & 2.4e-18 & \cite{Lange2021}\\
NPL-E3Yb+3    & \ce{^{171}Yb+}(E3) & 2.2e-18   & 2.5e-18 & \cite{Tofful2024}\\
PTB-Yb1E3     & \ce{^{171}Yb+}(E3) & 2.7e-18   & 2.4e-18 & \cite{Sanner2019}\\
IT-Yb1      & \ce{^{171}Yb} & 2.0e-17   & 2.7e-18 & \cite{Goti2023}\\
NPL-Sr1     & \ce{^{87}Sr} & 1.4e-17   & 2.7e-18 & \cite{Hobson2020}\\
PTB-Sr3     & \ce{^{87}Sr} & {---}   & 2.4e-18 & \cite{Schwarz2022}\\
\bottomrule
\end{tabular}
\label{tab:clocks}
\end{table}

In Table \ref{Tab:FrequencyRatios} we present 11 optical frequency ratio measurements between 6 optical clocks. The reported uncertainty is the total estimated uncertainty combining the systematic contribution of each clock, the statistical uncertainty and the uncertainty of the relativistic redshift \cite{Riedel2020,Denker2018}, which we note may differ between local and remote ratios. The optical fiber links contribute negligibly to the final uncertainties.

\begin{table}
\caption{Frequency ratios measured in this work, shown with the estimated uncertainties for
each measurement. See main text for discussion on the consistency of the results.}
\label{Tab:FrequencyRatios}
\footnotesize
\begin{tabular}{@{}{l}{l}{l}{l}{l}}
\toprule
No. & Clock 1    & Clock 2 & Value of frequency ratio & Fractional uncertainty\\
\midrule
1 &	SYRTE-Hg &	NPL-E3Yb+3  &	 1.757 572 821 468 312 99(10) &	\num{5.8e-17}\\
2 &	SYRTE-Hg &	PTB-Yb1E3  &	 1.757 572 821 468 313 043(99) &	\num{5.6e-17}\\
3 &	SYRTE-Hg &	IT-Yb1   &	 2.177 473 194 134 564 73(13) &	\num{6.1e-17}\\
4 &	SYRTE-Hg &	NPL-Sr1  &	 2.629 314 209 898 909 08(16) &	\num{6.1e-17}\\
5 &	PTB-Yb1E2 &	PTB-Yb1E3  &	 1.072 007 373 634 205 480(29) &	\num{2.7e-17}\\
6 &	NPL-E3Yb+3 &	PTB-Yb1E3  &	 0.999 999 999 999 999 992 9(77) &	\num{7.7e-18}\\
7 &	NPL-E3Yb+3 &	IT-Yb1  &	 1.238 909 231 832 259 570(30) &	\num{2.4e-17}\\
8 &	NPL-E3Yb+3 &	NPL-Sr1  &	 1.495 991 618 544 900 563(23) &	\num{1.5e-17}\\
9 &	PTB-Yb1E3 &	IT-Yb1  &	 1.238 909 231 832 259 569(27) &	\num{2.2e-17}\\
10 &	PTB-Yb1E3 &	NPL-Sr1  &	 1.495 991 618 544 900 560(22) &	\num{1.5e-17}\\
11 &	IT-Yb1 &	NPL-Sr1    &	 1.207 507 039 343 337 718(32) &	\num{2.7e-17}\\

\bottomrule
\end{tabular}
\end{table}

\begin{figure}
    \centering
    
        \includegraphics[width=0.72\textwidth]{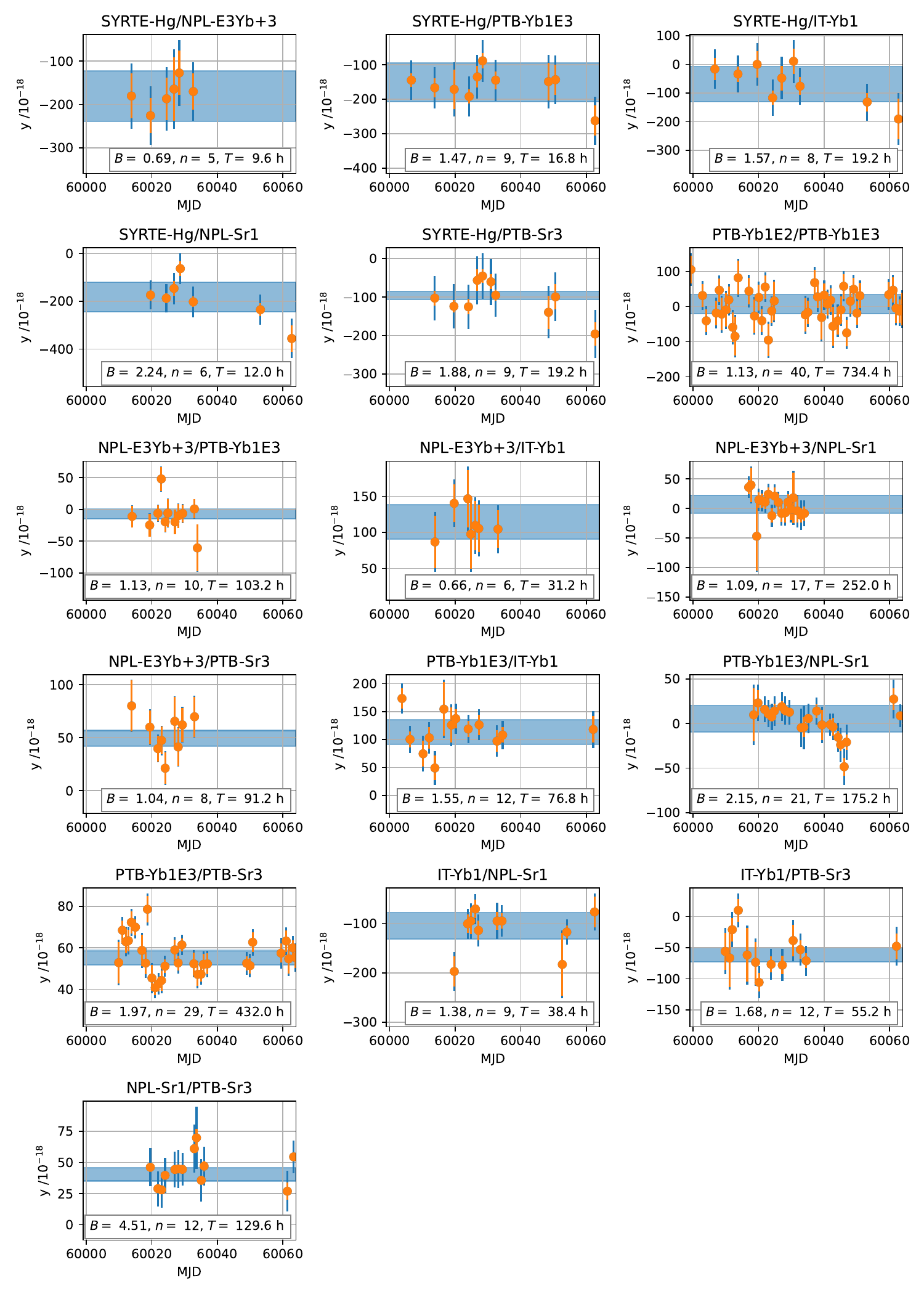}
    \caption{\footnotesize{Measured daily average frequency ratios between different clocks in this campaign. 
    Each optical frequency ratio is reported as a fractional offset, $y = R/R_0 - 1$, where $R$ is the measured value and $R_0$ is the corresponding reference frequency ratio obtained from a least-squares adjustment of standard frequencies (Appendix B of Ref.~\cite{Margolis2024}). The blue error bars represent the total uncertainties, evaluated as the sum in quadrature of statistical and systematic contributions. For panels involving PTB-Sr3, the systematic uncertainty of PTB-Sr3 is not included. The orange error bars represent the statistical uncertainties of each point. The box in each plot reports the Birge ratio $B$, the number of degrees of freedom $n$ and the total measurement time $T$. The blue regions show the average value with the total evaluated uncertainty,  with the statistical contribution inflated by the Birge ratio.}}
    \label{fig:scatter}
\end{figure}

\begin{figure}
    \centering

        \includegraphics[width=0.85\textwidth]{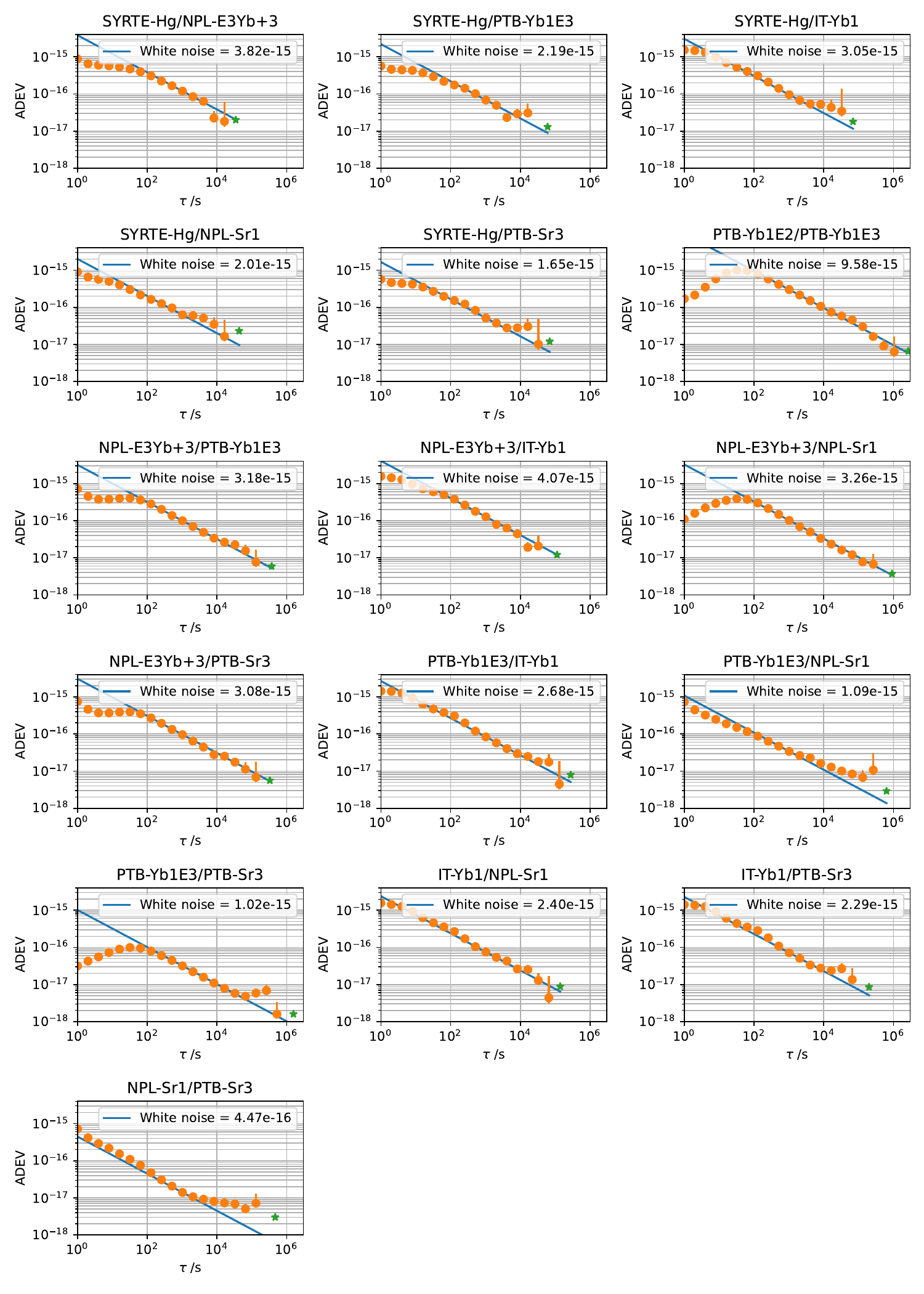}
    
    \caption{ Allan deviations (ADEV) for each of the measured frequency ratios in the campaign as a function of averaging time $\tau$. 
    The blue lines show fits assuming white frequency noise for each ratio (instability scaling as $\tau^{-1/2}$), with the legend reporting the value at \SI{1}{s}.
    The green stars show the statistical uncertainty of each ratio inflated by the Birge ratio if greater than 1, at the total measurement time. Error bars show the $1\sigma$ uncertainty of the Allan-deviation estimates, assuming white frequency noise~\cite{Stein1985}.}
    \label{fig:allan}
\end{figure}

The data analysis is similar to previous experiments (see \cite{Rocit2025,Clivati2022a} for details). 
For each measurement,  the statistical uncertainty was inflated by the Birge ratio (square root of the reduced chi-squared) calculated from the averages in daily bins, only if the Birge ratio was larger than 1. We observed Birge ratios ranging from 0.7 to 4.5, indicating some excess of daily scatter when statistical uncertainties are calculated from only the short-term white-frequency noise observed.   Other clock comparison measurements reported similar Birge ratios \cite{Rocit2025, Beloy2021, Dorscher2021, Ushijima2015, Ohmae2020}.
Figures \ref{fig:scatter} and \ref{fig:allan} show the measured frequency ratios averaged in daily bins and the corresponding Allan deviations.  We note that these Birge values and the departure of Allan deviations from pure white frequency noise at long averaging times could be related to measurements being carried out in different conditions or by systematic corrections applied daily. Nonetheless, unrecognized variations in systematic effects may also contribute.  Short-averaging-time deviations from white frequency noise in Fig. \ref{fig:allan} arise instead from the different servo time constants of the clocks \cite{Peik2005}, which range from approximately a few seconds to \SI{30}{s}.

The statistical uncertainties achieved by the end of the campaign range from approximately \num{1e-17} to  \num{2e-18}, as seen in Figure \ref{fig:allan}. The statistical uncertainties are typically negligible compared to the systematic uncertainty, with the exception of the frequency ratio NPL-E3Yb+3/PTB-Yb1E3.
Moreover, the instability of the ratio between NPL-Sr1 and PTB-Sr3 shows that the comparison by optical links does not limit the instability assessment at a level of \num{7e-16} at 1 s averaging time and $\num{5e-16} (\tau /\si{\second})^{-1/2}$   for $\tau>\SI{1}{s}$, where $\tau$ is the averaging time.

The uncertainty contributions from the optical frequency combs and fiber links are negligible at the level of the present measurements.
The stabilized fiber links used for optical-frequency transfer show instabilities below \num{1e-15} at \SI{1}{s}, reaching below \num{1e-18} after a few thousand seconds of measurement time, with characterized frequency offsets below the \num{1e-18} level~\cite{Clivati2022a, Cantin2021, Lisdat2016}.
Optical frequency combs used for optical-to-optical comparisons reach instabilities below \num{5e-16} at \SI{1}{s} and below \num{1e-19} after a few thousand seconds~\cite{Johnson2015, Barbieri2019, Nicolodi2014}, with state-of-the-art systems reaching below \num{1e-17} at \SI{1}{s} and below \num{1e-21} at \SI{e5}{s}~\cite{Benkler2019}.
Comb-related frequency offsets have also been characterized below the \num{1e-18} level.

\begin{figure}[h!]
    \centering
    
    \includegraphics[width=0.9\textwidth]{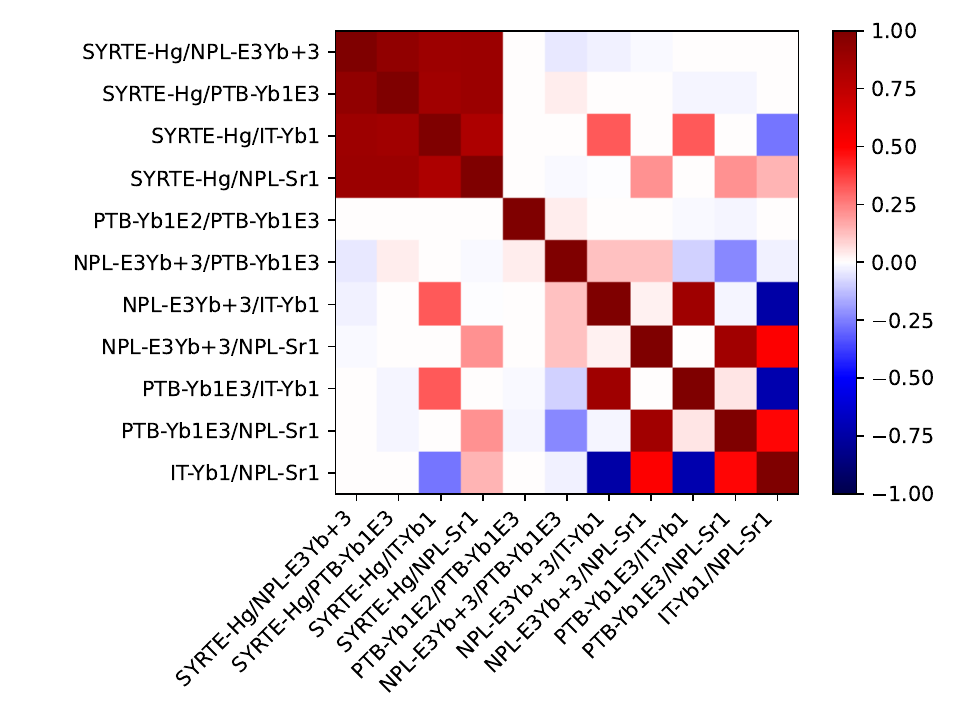}
    \caption{Graphical representation of the correlation coefficients between measurements in Table \ref{Tab:FrequencyRatios}. Numerical values of the correlation coefficients are reported in Appendix \ref{sec:corr}. }
    \label{fig:correlations}
\end{figure}

For multivariate measurements, it is important to report the correlation coefficients between the frequency ratio measurements, since the measurements depend on common quantities \cite{Rocit2025,ROCIT_D3}. For example, if two ratios involve a common clock, then the systematic shifts and statistics associated with that clock will introduce a correlation between the two frequency ratios. Appendix \ref{sec:corr} reports the correlation coefficients between the measurements in Table~\ref{Tab:FrequencyRatios}, calculated from the systematic and statistical uncertainties, including the uncertainty of the relativistic redshift.  Figure \ref{fig:correlations} shows a graphical representation. For 11 measurements, there are 55 correlation coefficients. Of these, 40  are non-zero ($\lvert r \lvert> 0.001$), with 11 being $\lvert r \lvert>0.5$. This indicates interdependence among the measured values, mainly caused by the systematic uncertainties of clocks common to different ratios \cite{Rocit2025,ROCIT_D3}, that should be considered when combining the different results \cite{Margolis2024}.  This is particularly pronounced for the ratios involving SYRTE-Hg, IT-Yb1, and NPL-Sr1 because the systematic uncertainties of those clocks are the dominant contribution to the total uncertainties.

\section{Discussion}
To discuss the results, we have plotted the frequency ratios involving different atomic species and compared them with previous measurements (carried out locally or via fibre link) in Figures \ref{fig:hist_ybplus}, \ref{fig:hist_hg}, and \ref{fig:hist_yb}. No prior measurements exist for the \ce{^{199}Hg}/\ce{^{171}Yb+}(E3) frequency ratio. Nevertheless, all ratios can be compared to the reference values provided in Appendix B of Ref. \cite{Margolis2024}, which result from a least-squares adjustment conducted in 2021, similar to the procedure used for determining the recommended values of physical constants \cite{Tiesinga2021}.
The uncertainties associated with these reference values are shown in the figures as gray bars and are derived from the self-consistent procedure using historical measurements, including past absolute frequency measurements and past optical frequency ratio measurements \footnote{A new least-square adjustment has been carried out in 2025 \cite{CCTF2025,BIPM-List}.  }.

\subsection{Comparison of \ce{^{171}Yb+}(E3) clocks}
The two \ce{^{171}Yb+}(E3) ion clocks at NPL and PTB are found to agree with each other at the level of \num{7.7e-18}, limited by the statistical uncertainty of the comparison. 
This improves upon the agreement between the two clocks with an uncertainty of \num{2.8e-16} observed in the ROCIT 2022 campaign level via Global Navigation Satellite System (GNSS) comparison, making this measurement the first international verification of two independently built optical clocks at a level below \num{1e-17}.

\begin{figure}[h!]
    \includegraphics[width=0.7\textwidth]{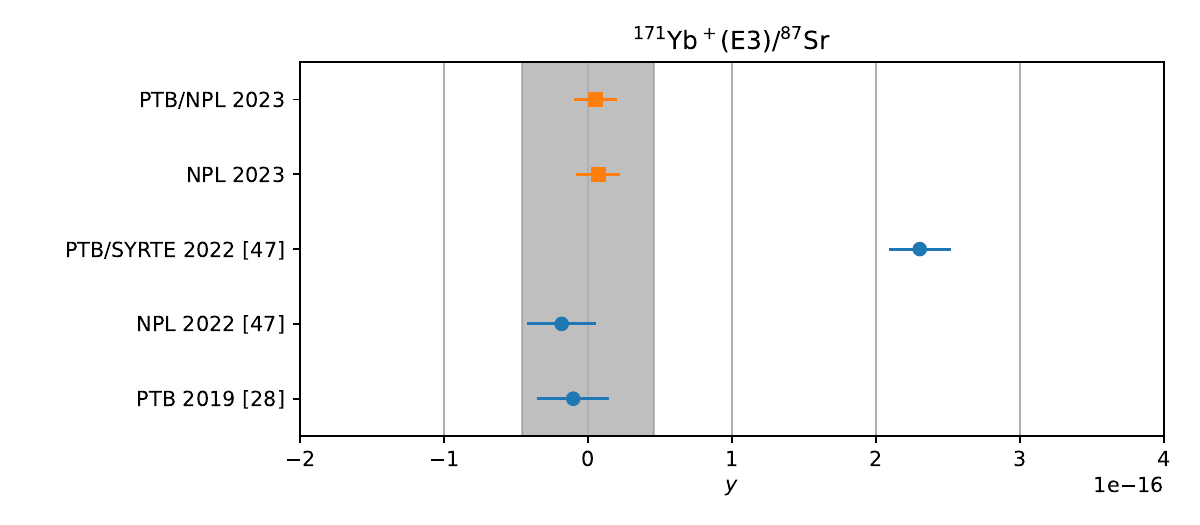}\\
    \caption{Comparison of the frequency ratios  \ce{^{171}Yb+}/\ce{^{87}Sr} measured in this campaign (orange squares) with other local or fibre-link measurements (blue circles).  Previous values are from Refs. \cite{Rocit2025, Dorscher2021}. Labels specify the institutes operating the \ce{^{171}Yb+} (E3) ion clock and the \ce{^{87}Sr} lattice clock, respectively, and the year of measurement. The optical frequency ratios are reported as fractional offsets, $y = R/R_0 - 1$, where $R$ is the measured value and $R_0$ is a reference frequency ratio obtained from a least-squares adjustment of standard frequencies (Appendix B of Ref.~\cite{Margolis2024}). The uncertainty of the reference value is indicated by the gray bar.}
    \label{fig:hist_ybplus}
\end{figure}

The agreement of the two \ce{^{171}Yb+}(E3) ion clocks at NPL and PTB can also be observed in the ratios with the \ce{^{87}Sr} lattice clock at NPL (see Fig. \ref{fig:hist_ybplus}). The measured \ce{^{171}Yb+}(E3)/\ce{^{87}Sr} ratios agree and have a lower uncertainty than previous local measurements at NPL and PTB \cite{Rocit2025, Dorscher2021}, as well as with earlier measurements via satellites techniques \cite{Riedel2020}. A discrepancy is observed with measurements involving SYRTE-Sr2 in the ROCIT 2022 campaign \cite{Rocit2025}.

\subsection{Frequency ratios with the \ce{^{199}Hg} clock}
\begin{figure}[h!]

    \includegraphics[width=0.7\textwidth]{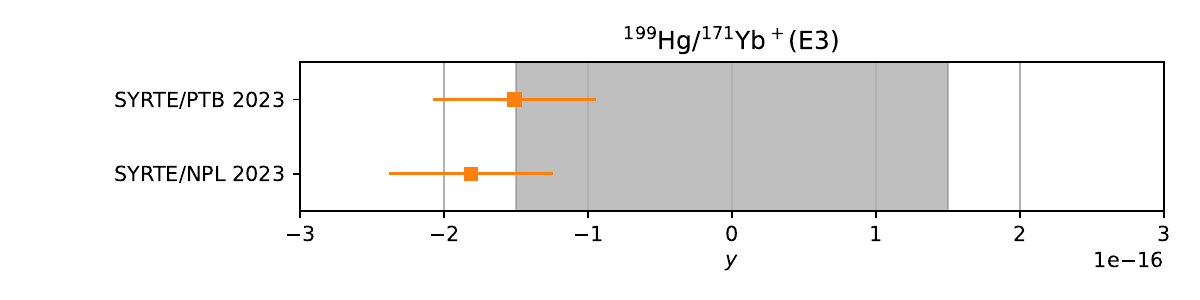}\\
    \includegraphics[width=0.7\textwidth]{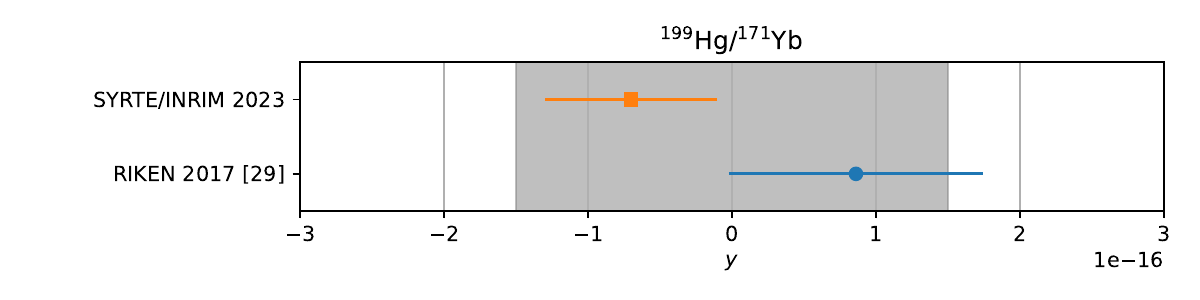}\\
    \includegraphics[width=0.7\textwidth]{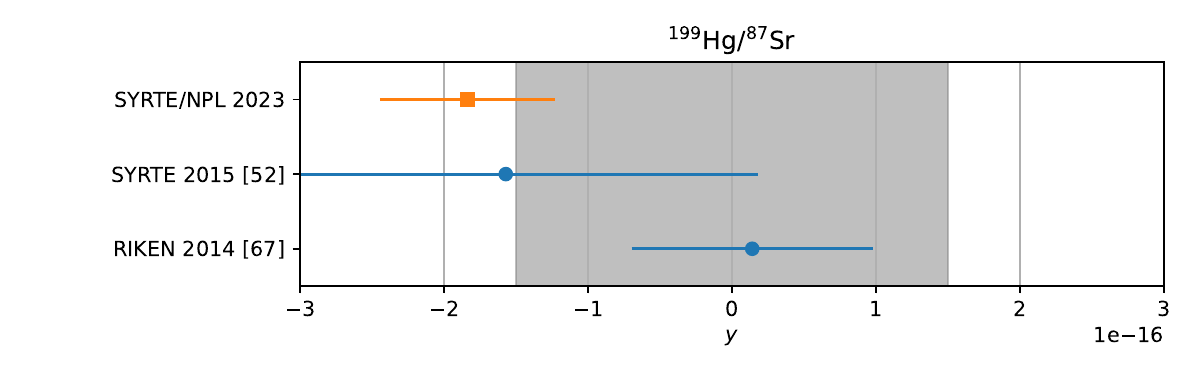}\\
    \caption{Comparison of the frequency ratios involving \ce{^{199}Hg} measured in this campaign (orange squares) with previous local measurements  (blue circles).  Previous values are from Refs. \cite{Ohmae2020, Yamanaka2015, Tyumenev2016}. Labels specify the institutes operating the clocks and the year of measurement. The optical frequency ratios are reported as fractional offsets, $y = R/R_0 - 1$, where $R$ is the measured value and $R_0$ is a reference frequency ratio obtained from a least-squares adjustment of standard frequencies (Appendix B of Ref.~\cite{Margolis2024}). The uncertainty of each reference value is indicated by a gray bar.}
    \label{fig:hist_hg}
\end{figure}

This campaign allowed us to measure four optical frequency ratios involving \ce{^{199}Hg}, as reported in Figure \ref{fig:hist_hg}. The frequency ratios between SYRTE-Hg and the \ce{^{171}Yb+} (E3) clocks at NPL and PTB agree at the 1$\sigma$ level with the reference frequency ratio (Appendix B of \cite{Margolis2024}) and have lower uncertainty. No previous direct measurements of this ratio have been published.

Similarly, the frequency ratio between SYRTE-Hg and IT-Yb1 agrees well with—and has a lower uncertainty than—the reference frequency ratio, as well as with the only previous direct measurement of the \ce{^{199}Hg}/\ce{^{171}Yb} frequency ratio at RIKEN \cite{Ohmae2020}.   

The frequency ratio between SYRTE-Hg and the NPL-Sr1 is consistent within 2$\sigma$ of the two previous measurements of the \ce{^{199}Hg}/\ce{^{87}Sr} frequency ratio at SYRTE \cite{Tyumenev2016} and RIKEN  \cite{Yamanaka2015}, as well as with the reference frequency ratio.

For completeness, we note that the SYRTE-Hg clock participated in 4 clock comparison campaigns between 2018 and 2020, operating in a similar manner to the 2023 campaign \cite{Zyskind2024}. Some inconsistencies were observed between those earlier results and Ref. \cite{Yamanaka2015}, at a typical maximum level of \num{2.5e-16} (4$\sigma$), with no definitive source identified. In contrast, the 2023 campaign provides a significantly more reliable dataset for the SYRTE-Hg clock, owing to its extended duration and enhanced control of systematic effects. In particular, a revision of several systematic effects—such as the quadratic Zeeman shift, the cold collisional shift, and the background gas collisional shift—has contributed to increased confidence in these results.




\subsection{Frequency ratios with the \ce{^{171}Yb} clock}
\begin{figure}[h!]
    \includegraphics[width=0.7\textwidth]{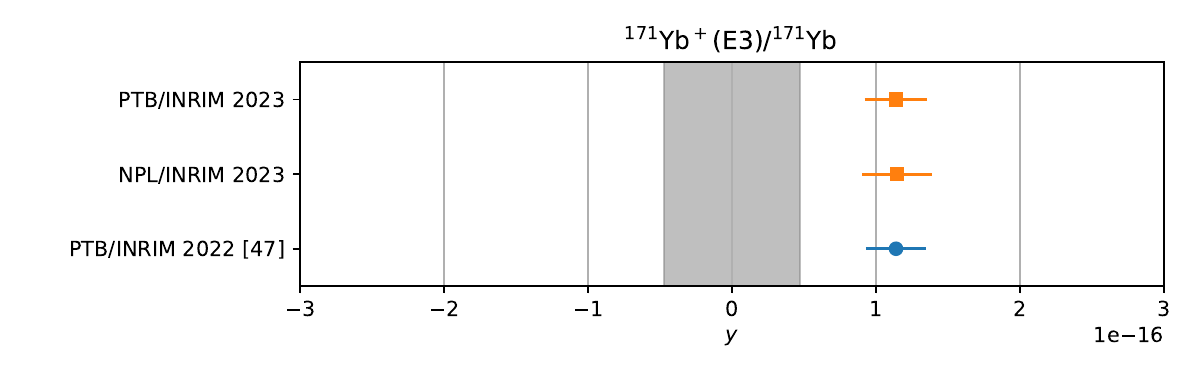}\\
    \includegraphics[width=0.7\textwidth]{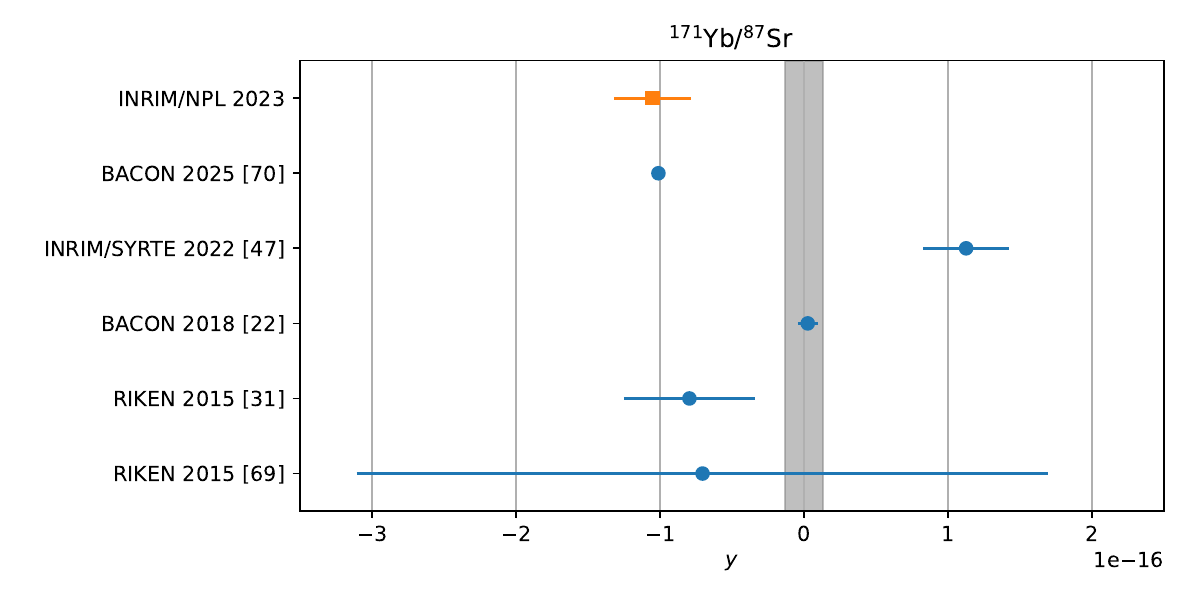}
    \caption{Comparison of the frequency ratios involving \ce{^{171}Yb} measured in this campaign (orange squares) with other local or fibre measurements (blue circles).  Other values are from Refs. \cite{Rocit2025, Nemitz2016,Takamoto2015,Beloy2021,Aeppli2025}. Labels specify the institutes operating the clocks and the year of measurement. For the ratio \ce{^{171}Yb}/\ce{^{87}Sr} only measurements with uncertainty $< \num{3e-16}$ are shown. The optical frequency ratios are reported as fractional offsets, $y = R/R_0 - 1$, where $R$ is the measured value and $R_0$ is a reference frequency ratio obtained from a least-squares adjustment of standard frequencies (Appendix B of Ref.~\cite{Margolis2024}). The uncertainty of each reference value is indicated by a gray bar.}
    \label{fig:hist_yb}
\end{figure}%

The frequency ratios between IT-Yb1 and the \ce{^{171}Yb+}(E3) clocks (NPL-E3Yb+3 and PTB-Yb1E3) in this campaign agree with the ratio measured in the ROCIT 2022 campaign between INRIM and PTB (see Fig. \ref{fig:hist_yb}). The difference is explained by the statistical uncertainty alone (\num{1e-17}) indicating a good reproducibility of the clocks. 
These values differ from the reference frequency ratio by 2$\sigma$ of the combined uncertainty of \num{5e-17} but we note that no direct measurement of this frequency ratio was used for the adjustment \cite{Margolis2024}, which was instead driven by the other frequency ratios, including \ce{^{171}Yb}/\ce{^{87}Sr}.
Moreover, these frequency ratios also agree with the GNSS measurements between PTB, NPL and NMIJ made during the ROCIT 2022 campaign, but there the uncertainty was an order of magnitude higher.

The frequency ratio between IT-Yb1 and  NPL-Sr1  disagrees by 4 $\sigma$  with the 2018 measurement of the \ce{^{171}Yb}/\ce{^{87}Sr} frequency ratio between NIST and JILA (BACON collaboration), in the USA \cite{Beloy2021} as well as an earlier measurement in the ROCIT 2022 campaign \cite{Rocit2025} (see Fig. \ref{fig:hist_yb}).  The ratio is consistent with the measurements at RIKEN \cite{Nemitz2016,Takamoto2015} as well as with a more recent measurement by the BACON collaboration \cite{Aeppli2025}.


\subsection{\ce{^{171}Yb+}(E2)/\ce{^{171}Yb+}(E3) frequency ratio}
The \ce{^{171}Yb+}(E2)/\ce{^{171}Yb+}(E3) frequency ratio measured locally at PTB agrees within a statistical uncertainty of \num{1e-17} alone with previous measurements obtained with the same clocks \cite{Lange2021,Rocit2025}.

\section{Conclusions}

We report new optical frequency ratio measurements between seven optical clocks from a two-month-long clock comparison campaign conducted in early 2023, involving four national metrology institutes in Europe connected by fiber links. Among those, we verified the ratio between the two \ce{^{171}Yb+}(E3) ion clocks at NPL and PTB  with an uncertainty of \num{7.7e-18}, which is the first international check of the consistency of optical clocks at this level.  The \ce{^{199}Hg} clock at SYRTE was compared to the other clocks in the network, providing new direct frequency ratios with uncertainties at the \num{e-17} level. We confirmed the frequency ratio between the \ce{^{171}Yb} clock at INRIM and the \ce{^{171}Yb+}(E3) clock at PTB measured one year before in the ROCIT 2022 campaign, also at the \num{e-17} level.
Some frequency ratios measured in this campaign show inconsistencies with previous measurements, in particular among some \ce{^{171}Yb}/\ce{^{87}Sr} ratios, as also observed in a recent campaign conducted in the USA~\cite{Aeppli2025}.
However, no experimental issue or overlooked systematic effect has so far been reported for any of these measurements.

Resolving these inconsistencies is a challenge towards the redefinition of the second and it will require further effort from the metrology community worldwide. 
This and future comparison campaigns, together with global least-squares adjustments of clock comparison data, will provide important input for the critical analysis of recent experiments and may help reveal previously unidentified experimental issues.

The European network of optical fibers provides continued opportunities to compare clocks in different countries, with negligible contributions to the uncertainties from the links. Beyond the importance for metrology and fundamental physics, the establishment of clock  and fiber networks supports broader areas of research, for example  contributing to  more accurate geodetic models at regional and global scales \cite{Bondarescu2015,Lion2017,Mueller2017}, enabling quantum communication over long distances \cite{Clivati2022}, and facilitating frequency distribution to radio-telescope facilities \cite{Clivati2020a}.

\begin{acknowledgments}

We acknowledge experimental support from Daniele Nicolodi, Thomas Legero and Uwe Sterr by providing an ultrastable laser reference as local oscillator for the optical clocks at PTB. 

This work includes contributions from the EMPIR projects 20FUN08 NEXTLASERS and 20FUN01 TSCAC. These projects have received funding from the EMPIR programme co-financed by Participating States and from the European Union’s Horizon 2020 research and innovation programme.
This work was partially supported by the project 22IEM01 TOCK. This project has received funding from the European Partnership on Metrology, co-financed from the European Union’s Horizon Europe Research and Innovation Programme and by the Participating States.
Optical fiber links from the French REFIMEVE network are supported by Program “Investissements d’Avenir” launched by the French Government and implemented by Agence Nationale de la Recherche with references ANR-11-EQPX-0039 (Equipex REFIMEVE+) and ANR-21-ESRE-0029 (ESR/Equipex+T-REFIMEVE).
M. Mazouth-Laurol received funding from the European Union’s Horizon 2020 research and innovation programme under Grant Agreement No. 951886 (CLONETS-DS).
M. Tønnes received funding from DIM SIRTEQ network of Conseil Régional Île de France (ATH-2019 ONSEPA).
C. Zyskind received funding from Sorbonne Université (PhD grant CMA FQPS n°ANR-21-CMAQ-0001 dans le cadre de France 2030).
NPL acknowledges support from the Department for Science, Innovation and Technology - National Measurement System programme.
PTB acknowledges support by the Deutsche Forschungsgemeinschaft (DFG, German Research Foundation) under Germany’s Excellence Strategy – EXC-2123 QuantumFrontiers – Project-ID 3908379.67 and SFB 1464 TerraQ – Project-ID 434617780 – within projects A04 and A05, SFB 1227 DQ-mat – Project-ID 274200144, within Project B02.
This work was partially supported by the Max Planck–RIKEN–PTB Center for Time, Constants and Fundamental Symmetries.

\end{acknowledgments}

\section*{Data availability}
The data that support the findings of this article are openly available \cite{data_on_zenodo}.

\appendix

\section{Correlation coefficients}
\label{sec:corr}
Table \ref{Tab:FibreCorr} reports the correlation coefficients between the measurements in Table \ref{Tab:FrequencyRatios} calculated from the systematic and statistical uncertainties, including the uncertainty of the relativistic redshift \cite{Rocit2025, ROCIT_D3}.

\begin{small}
\begin{longtable}{l@{$=$}S[table-format=+2.6] l@{$=$}S[table-format=+2.6] l@{$=$}S[table-format=+2.6] l@{$=$}S[table-format=+2.6]}

\caption{Correlation coefficients between  measurements in Table \ref{Tab:FrequencyRatios}.} \label{Tab:FibreCorr}\\

\toprule
\endfirsthead

\multicolumn{8}{c}%
{{\tablename\ \thetable{} (continued)}} \\
\toprule
\endhead

\bottomrule
\endfoot

\bottomrule
\endlastfoot
$r(\text{1,2})$ & 0.925	& $r(\text{1,3})$ & 0.882	& $r(\text{1,4})$ & 0.887	& $r(\text{1,6})$ & -0.045\\
$r(\text{1,7})$ & -0.027	& $r(\text{1,8})$ & -0.013	& $r(\text{1,10})$ & 0.007	& $r(\text{1,11})$ & 0.004\\
$r(\text{2,3})$ & 0.864	& $r(\text{2,4})$ & 0.883	& $r(\text{2,5})$ & 0.005	& $r(\text{2,6})$ & 0.038\\
$r(\text{2,9})$ & -0.022	& $r(\text{2,10})$ & -0.020	& $r(\text{3,4})$ & 0.814	& $r(\text{3,7})$ & 0.322\\
$r(\text{3,9})$ & 0.320	& $r(\text{3,11})$ & -0.269	& $r(\text{4,6})$ & -0.013	& $r(\text{4,7})$ & -0.005\\
$r(\text{4,8})$ & 0.219	& $r(\text{4,10})$ & 0.212	& $r(\text{4,11})$ & 0.143	& $r(\text{5,6})$ & 0.037\\
$r(\text{5,9})$ & -0.015	& $r(\text{5,10})$ & -0.022	& $r(\text{6,7})$ & 0.120	& $r(\text{6,8})$ & 0.117\\
$r(\text{6,9})$ & -0.092	& $r(\text{6,10})$ & -0.234	& $r(\text{6,11})$ & -0.029	& $r(\text{7,8})$ & 0.028\\
$r(\text{7,9})$ & 0.867	& $r(\text{7,10})$ & -0.018	& $r(\text{7,11})$ & -0.750	& $r(\text{8,10})$ & 0.866\\
$r(\text{8,11})$ & 0.506	& $r(\text{9,10})$ & 0.049	& $r(\text{9,11})$ & -0.720	& $r(\text{10,11})$ & 0.491\\

\end{longtable}
\end{small}

\bibliography{march2023}

\end{document}